\documentclass[12pt]{iopart}
\usepackage[table,xcdraw]{xcolor}
\usepackage{amsfonts}
\usepackage{amssymb}
\usepackage{booktabs}
\usepackage{algorithmic}
\usepackage{algorithm}
\usepackage{graphicx}
\usepackage{textcomp}
\usepackage[bottom=1in,right=1.05in]{geometry}
\usepackage{amsthm}
\usepackage{subfig}

\newtheorem{remark}{Remark}

\begin{document}

\title[]{Information Transfer Rate in BCIs: Towards Tightly Integrated Symbiosis}

\author{Suayb S. Arslan and Pawan Sinha}

\address{Department of Brain and Cognitive Sciences, \\ Massachusetts Institute of Technology,  \\ Cambridge, MA, USA, 02139.}
\ead{sarslan@mit.edu, psinha@mit.edu}
\vspace{10pt}

\begin{abstract} \\
\textit{Objective.} The information transmission rate (ITR), or effective bit rate, is a popular and widely used information measurement metric, particularly popularized for SSVEP-based Brain-Computer (BCI) interfaces. By combining speed and accuracy into a single-valued parameter, this metric aids in the evaluation and comparison of various target identification algorithms across different BCI communities. In order to calculate ITR, it is customary to assume a uniform input distribution and an oversimplified channel model that is memoryless, stationary, and symmetrical in nature with discrete alphabet sizes. To accurately depict performance and inspire an end-to-end design for futuristic BCI designs, a more thorough examination and definition of ITR is therefore required. \textit{Approach.} We model the symbiotic communication medium, hosted by the retinogeniculate visual pathway,  as a discrete memoryless channel and use the modified capacity expressions to redefine the ITR. We \textcolor{black}{leverage a result for directed graphs} to characterize the relationship between the asymmetry of the transition statistics and the ITR gain \textcolor{black}{due to} the new definition, leading to potential bounds on data rate performance. \textit{Main Results.} On two well-known SSVEP datasets, we compared two cutting-edge target identification methods. Results indicate that the induced DM channel asymmetry has a greater impact on the actual perceived ITR than the change in input distribution. Moreover, it is demonstrated that the ITR gain under the new definition is inversely correlated with the asymmetry in the channel transition statistics. Individual input customizations are further shown to yield perceived ITR performance improvements. Finally, an algorithm is proposed to find the capacity of binary classification and further discussions are given to extend such results to multi-class case through ensemble techniques. \textit{Significance.} We anticipate that the results of our study will contribute to the characterization of the highly dynamic BCI channel capacities, performance thresholds, and improved BCI stimulus designs for a tighter symbiosis between the human brain and computer systems while ensuring the efficient utilization of the underlying communication resources.
\end{abstract}

%
%
%
%
%

\section{Introduction}
\label{sec:introduction}

The primary goal of brain-computer interfaces (BCIs) is to provide new channel formations for communication and control between the human brain and its surrounding devices with computational capabilities \cite{wolpaw2002}. One of the most commonly used modality in BCIs is known as Electroencephalography (EEG), which results due to stimuli generating electrical fields caused by synchronously firing neuron populations in the human brain. The majority of BCI research efforts are focused on developing effective stimuli, experimental protocol developments and target identification/classification algorithms to boost information transfer and eventually help novel communication paradigms to emerge such as found in recent semantic communications \cite{Geuze2014}. A summary of the system diagram for a cannonical BCI is shown in Fig. 1, where the components of the system and their interactions are illustrated. Despite low information rates, different types of BCIs are actively employed in various applications nowadays, ranging from clinical deployments \cite{Mak2009} to entertainment world, including but not limited to gaming  \cite{Nij2008}, \cite{Ertekin2022} and virtual/augmented reality \cite{Putze2019}. Such commonplace applications has become quite encouraging and pushed current state-of-the-art BCI research forward, enabling tight coupling and enhanced human involvements in hybrid (human and machine combined) systems \cite{Branson2014}.  

\begin{figure}[!b]
\centerline{\includegraphics[width=0.9\columnwidth]{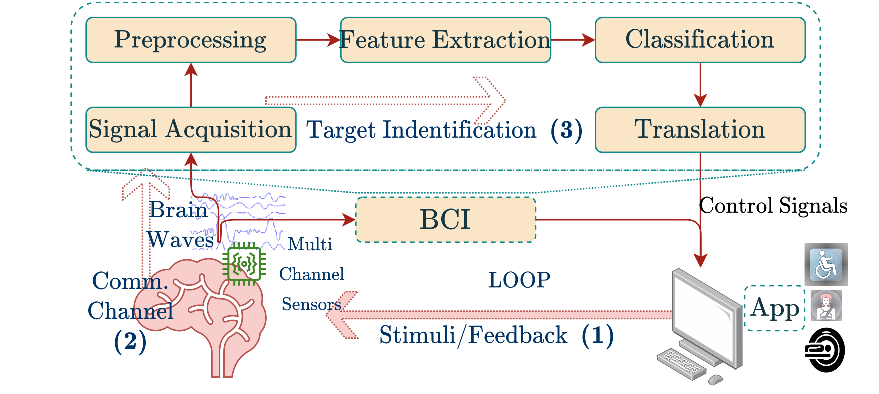}}
\caption{Closed loop system of generic BCIs. A  BCI system consists of (1) the stimulation/feedback, (2) the communication channel and (3) the target identification/classification.}
\label{fig1}
\end{figure}

As shown in Fig. 1, a generic BCI system typically consists of three main parts: (1) the stimulation/feedback generation, (2) the communication channel hosted by the participating subject, and (3) the target identification system. To be able to deliver fast communication rates and form a close symbiosis between the human brain and computing devices, these components must work in concert. In other words, a joint design of these components is required to modulate and control the communication channel formed by the combined effects of these three components. On the other hand, for a truly coupled system design in which the stimulus generation and TIs co-adapt to each other \cite{verhoeven2015}, better understanding of the performance evaluations, controlling measures and the underlying statistical channel formed is needed. Once established, new research directions can be explored such as optimal stimuli designs, developing useful performance bounds \textcolor{black}{on the data transfer rates} and \textcolor{black}{develop genuine interfaces} for enhanced end-user experience \textcolor{black}{\cite{diva2019}}.

\subsection{\textcolor{black}{Performance Evaluation}}

Assessments of BCI system performance are typically performed at two levels of evaluation, namely \textit{user-level} and \textit{system-level}. User\textcolor{black}{-level} performance is measured by the degree of congruence between user intent and the signal feature(s) the BCI uses to identify the intent. The user-level quality control heavily depends on the visual setup, the selection and presentation of stimuli, and how the stimulation is carried out (usually forming a protocol).  However, system-level performance evaluations are often done in terms of target identification speed and classification accuracy, conventionally tied to each other. Fair comparisons are difficult since these two criteria, when articulated independently, are affected by the program's capability as well as how well the system combines the user's control with that application. 
\textcolor{black}{Additionally, it can be challenging to theoretically quantify the structural representations of content since they may be influenced by perception \cite{Lungarella2005}. In
many applications, the interpretation of quality may differ across different users, leading to indispensable
subjectivity. It is anticipated that forthcoming BCI systems will impose less cognitive burden, as stimulation paradigms become more personalized and rely on individualized internal representations \cite{Xu2021}. As a result, many researchers have opted to use context-sensitive definitions of information instead \cite{Thorton2013}. However, in mainstream BCI applications\footnote{\textcolor{black}{In some of these applications, there may be a requirement for enhanced definitions of information that consider the contextual dependence in achieving data transfer rates. Nonetheless, this
would pose additional challenges by introducing greater subjectivity in evaluating performance across
various BCIs.}} where systematic and well-defined stimulation paradigms and protocols are used, the semantic meanings of inputs and outputs are established beforehand, and users are essentially trained on the internal structure of the presented symbols that are to be communicated. This is what fundamentally makes objective (context-independent) definitions of information still relevant for practical BCI implementations.}
Information transfer rate (ITR), cited primarily in \cite{McFarland2012} and \cite{sahar2019}, is one of a number of \textcolor{black}{context-independent information} measuring tools developed in response to the demand for a single theoretical measure that combines accuracy and speed. \textcolor{black}{In our work,} information transfer is quantified according to information theory measures such as entropy, pioneered by Shannon's seminal work in 1948 \cite{shannon1948}.

As shall be demonstrated, it is not hard to realize that ITR can be used to measure the magnitude of coupling in a BCI communication setting as well as the levels of attention and consciousness, which can be leveraged heavily in passive-BCI settings as well \cite{cotrina2014}. In the past, ITR is employed in BCIs both using P300 paradigm \cite{DeVos2014} and in particular steady-state
visual evoked potentials (SSVEPs) due to proven high data communication performance. For instance, recent progression towards task-related component analysis for SSVEPs is shown to achieve rates up to 325 bits/min ITR in a cue-guided 40-character spelling task \cite{Nakanishi2018}. \textcolor{black}{For that reason, the rest of our discussions will rely on SSVEP-based BCIs due to their excellent information transfer rates. We primarily note that } such performance has been possible due to carefully designed stimulus generation and Target Identification (TI) techniques. Stimuli generation involves embedding information into frequencies and phases of the signals (neural modulation) in an SSVEP paradigm \cite{JFPM2014}.  TI  requires compensating for the noise and degradations imposed on the information passing through a channel induced by the BCI, resting on the physical medium of the \textit{retinogeniculate} visual pathway. It turns out that many information processing strategies are used throughout the visual system and the basic approach is to lump them into a coarse description of an information link \cite{Essen1995}. The main objective of the majority of the past work has been to maximize the information transfer rate that can be transmitted over this induced channel \cite{hanx2019}. Although the BCI channel properties are determined by the choice of stimulation  and the target identification methods combined, without the accurate measurement of the ITR performance, it would not be possible to assess the user experience, develop successful stimuli design techniques and conduct joint end-to-end optimizations of BCIs.


Preconditions and deficiencies of the conventional ITR definition are considered in a number of past studies. For instance, \cite{yuan2013} summarizes the problems with the conventional definition and particularly emphasizes the significance of the channel parameter estimations (accuracy or false alarm rates) in online synchronous BCIs. The same study proposed a task-oriented online BCI test in the hope to help with the real-world applications.
Moreover, recent works such as \cite{daSilva2020} proposed to use an alternative closed-form formulation derived in  \cite{Muroga1953} for the ITR computation via making approximations such as removing negativity constraint on the input distribution using only a fairly limited dataset. In fact, this closed form's usefulness is restricted to  square and non-singular channel transition characterizations \cite{arimoto1972}. Moreover, no further analysis or intuition is presented in both studies in terms of the channel transition characteristics, the stimuli design, and means and strategies for achieving the capacity of the underlying BCI channel \textcolor{black}{models}. 

\subsection{\textcolor{black}{Contributions and Organization}}

One of the main objectives of this study is to outline the basic principles of conventional ITR definition and in order to re-express its deficiencies, highlight the challenges of channel characterization problems between the human brain and the computer system. \textcolor{black}{As part of our contributions,} we realize that the performance characterization of any technique for an asymmetric and non-stationary channel requires a \textcolor{black}{precise} computation procedure to accurately determine the practical ITR the subjects experience. \textcolor{black}{Moreover, we outline a numerical calculation of the ITR and provided a few} directions for tighter symbiosis within the context of joint system design. \textcolor{black}{More specifically,} we demonstrate that \textcolor{black}{our analysis} can be used to design better input \textcolor{black}{sequences as well as novel design of user interfaces} for \textcolor{black}{futuristic} BCIs to maximize the information flow in the induced communication channel. For instance, we have found that customizing the stimuli for each subject increases the channel capacity dramatically and \textcolor{black}{an end-to-end} joint system design seems to reduce the performance gap due to using different TIs \textcolor{black}{in SSVEP settings. Finally, we have provided a preliminary algorithm for finding fundamental data transfer rate limitations of a two-symbol BCI as a proof-of-concept.} 

The rest of the paper is organized as follows. In Section II, we introduce the generalized DM channel model and rephrased the conventional ITR definition. We report the deficiencies as well as workarounds through integrating the algorithmic capacity calculations into the ITR definition, subject to information theoretic bounds. In Section III, we present a few results to distinguish different ITR definitions, and explore the asymmetry in channel transition statistics and establish the ties with the \textcolor{black}{channel} conditional entropy. We discuss some of the important implications of our results in Section IV. For instance, we have shown a potential performance (ITR) upper bound using ensemble learning with binary weak learners. Finally, we conclude our paper in Section V by prodiving a few future directions.

\section{Methods}
\label{sec:ITR}

\subsection{Channel Model and Conventional ITR Definition}
Let us consider a discrete BCI communication system, where one of the $M$ symbols from the set $\mathcal{X}  = \{x_1,\dots,x_M\}$ is to be transferred at a given time. It is quite typical to express BCI system performance in terms of the information transfer rate (ITR) \cite{McFarland2012}. This is  expressed in bits per trial observation window $T$ \cite{wolpaw2002} as,
\begin{eqnarray}
    \texttt{ITR} = \log_2(M) + P(T)\log_2(P(T)) + (1-P(T))\log_2\left(\frac{1-P(T)}{M-1}\right)\label{eqn1ITR}
\end{eqnarray}
where $M$ is the number of targets and $P(T)$ is the aggregate average accuracy of the target identification algorithm. Note that the trial time dependency of the accuracy is crucial in this formulation. Equation (\ref{eqn1ITR}) is derived from the popular mutual information measure defined for two random variables $X$ and $Y$ as
\begin{eqnarray}
    I(X;Y) &=& H(Y) - H(Y|X) \\
    &=& \sum_{y \in \mathcal{Y}} P_Y(y) \log_2\left(\frac{1}{P_Y(y)}\right) - \sum_x P_X(x) H(Y | X = x)
\end{eqnarray}
where $H(.)$ is the entropy, $X \in \mathcal{X}$ represents the discrete source taking on one of the $M$ target classes and $Y \in \mathcal{Y}$ (typically $\mathcal{Y} = \mathcal{X}$) is the predicted output at the other end of the BCI system with distribution $P_Y(y)$. \textit{Capacity} ($\mathcal{C}$) is defined to be the supremum of the mutual information over all input (probability) distributions $P_X(x)$ \textcolor{black}{i.e.,
\begin{eqnarray}
    \mathcal{C} = \sup_{P_X(x)} I(X;Y)
\end{eqnarray}}
In the case of perfect communication ($P(T) \rightarrow 1$) the ITR will simply be $\log_2(M)$, the number of bits used to represent all targets assuming these targets have equal probability of occurring i.e., $1/M$ (uniform  input distribution i.e., $P_X(x) = \frac{1}{M}$) \cite{cover1999}. 

Equality \textcolor{black}{given for ITR} (\ref{eqn1ITR}) is based on the capacity of a symmetric Discrete Memoryless Channel  (DMC) that errs with an equal probability $\frac{1-p}{M-1}$ in favor of all other $M-1$ classes. In addition, $T$ is expressed in terms of seconds and the \texttt{ITR} is usually scaled with $60/T$ and expressed in terms of bits/min. Let us define the channel transition matrix of the induced DMC and express it as follows,
$$\mathbf{P} = \left[ \matrix{  p_{1,1} & p_{1,2} & \dots  & p_{1,M} \cr
p_{2,1} & p_{2,2} & \dots  & p_{2,M} \cr
\vdots & \vdots & \vdots & \ddots \cr
p_{M,1} & p_{M,2} & \dots  & p_{M,M} \cr} \right]$$
with $\sum_j p_{i,j} = 1$ for $i \in \{1,2,\dots,M\}$. We use the short-cut notation $P_{Y|X}(y_j|x_i) = p_{i,j}$ as each entry of $\mathbf{P}$ to refer to the channel transition/conditional probability distribution. Note that symmetry assumption in the channel transition statistics i.e., the distribution of the probability of erring over all non-target values make the uniform input distribution achieve the DMC capacity. Hence,
\begin{eqnarray}
    \max_{P_X(.)} I(X;Y) &= \left[ \log_2(M) - \sum_j p_{i,j} \log_2\left(\frac{1}{p_{i,j}}\right) \right] = \texttt{ITR}
\end{eqnarray}

Although we have assumed the possible number of outcomes to be $M$ i.e., $|\mathcal{Y}|=M$, the channel outputs can be expanded to include ``erasures" i.e., making no decision on the final output, leading to the channel transition matrix $\mathbf{P}$ to be of size $M \times (M+1)$. Both channel models  are illustrated in Fig. \ref{fig1}. 

\begin{figure}[!t]
\centerline{\includegraphics[width=0.9\columnwidth]{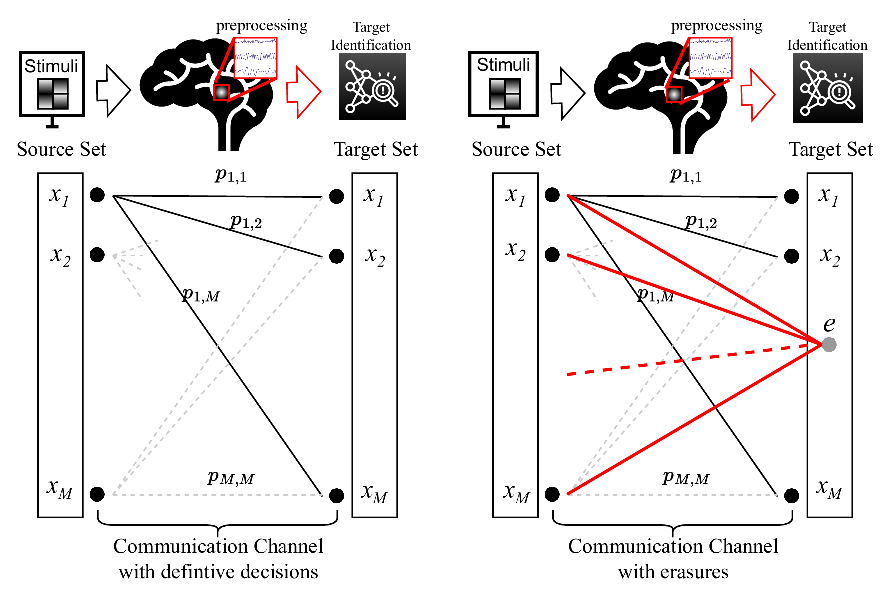}}
\caption{Typical discrete BCI Channel Models for symbol/character communications. A discrete set of characters are communicated. In case, the communication reliability is below a threshold, the communicated character can be assumed to be erased. In the figure, $\mathbf{e}$ is used to represent erasure. }
\label{fig1}
\end{figure}

\subsection{Deficiencies and Workarounds}

Some of the major deficiencies of the conventional ITR definition have already been articulated in \cite{yuan2013}. One of these deficiencies is the assumption that the input (the stimuli) has uniform distribution of $M$ \textcolor{black}{input} symbols. However, this assumption is not necessarily true and optimal \textcolor{black}{in realistic settings}. For instance, in a speller application, the characters of the English alphabet are not necessarily used equally frequently in everyday language and hence in an online experiment, some of the characters would naturally be intended to spell more often. In addition, transition probabilities of the underlying channel are not necessarily symmetric and stationary \cite{williams2018}. 
\textcolor{black}{In an SSVEP-based BCI, the ability to distinguish between two targets is determined by factors such as the frequency and phase selections, as well as the spatial distance at which these targets are presented to the subjects.}
Therefore, the assumption $p_{i,j}=\frac{1-p}{M-1}$ for all $i,j$ satisfying $i\not=j$ (i.e., equal probability transitions to all non-target classes) is not necessarily totally accurate. 

\textcolor{black}{Calculating the capacity of the dynamic channel model described in our work poses a challenge because the channel transition matrix is dependent not only on the stimulus design (how information is encoded into the visual pathway) but also on the methods used for identifying the target, which is quite untypical in classical digital communications.} In addition, as the observation window ($T$) widens, accuracy is reported to increase since the identification is performed via observation of more time series data \cite{cruz2013}. However, the observation window being larger will change the stochastic nature of the channel (and its parameters) i.e., the channel transition matrix indeed is a function of time. The extent of the introduced channel memory is yet another challenge to tackle. Thus, instead of considering the entire timeline, we consider specific time points such that the dependency of the channel transition matrix at those points is sufficiently eliminated. Some of the previous works proposed closed form expressions \cite{AshCapacity1965} under certain assumptions about $P_x(x)$ and non-singularity assumption for $\textbf{P}$, which is highly unlikely for larger window lengths. \textcolor{black}{Our approach in this study involves treating the channel as a DMC without any preconceived notions about its statistical nature. We first estimate the transition statistics and then use numerical methods to calculate the capacity. Finally, we compare our ITR results with those obtained using the conventional formulation.}


\subsection{\textcolor{black}{An iterative definition of ITR}}

\textcolor{black}{In this section, we have provided the details of the proposed ITR definition and its numerical computation, derived primarily from the capacity results of asymmetric DMCs. Let us begin with $M=2$ binary symbol communication case as the baseline which leads to a closed form expression.} 

\subsubsection{Exemplary Case: ``Binary Classification":}
Let us suppose $\mathcal{X} = \mathcal{Y} = \{x_1,x_2\}$ i.e., the input is one of the two possible symbols with $P_X(x_1)=p_x$. This would correspond to differentiation of two different classes such as face/non-face \cite{Zhang2012} or familiar/non-familiar (target/non-target paradigm) dichotomies. Thus, we can express the mutual information
\begin{eqnarray}
    I(X;Y) &=& H(Y) - H(Y|X) \\
    &=& h(p_x(1-p_{1,2})+(1-p_x)p_{2,1}) -p_xh(p_{1,2}) - (1-p_x)h(p_{2,1}) \label{eqninf}
\end{eqnarray}
where $h(x) = -x\log(x)-(1-x)\log(1-x)$ is the binary entropy function. Setting the derivative with respect to $p_x$ to zero, we obtain
\begin{eqnarray}
    \frac{1}{p_x(1-p_{1,2}-p_{2,1})+p_{2,1}} - 1 = 2^{\frac{h(p_{1,2})-h(p_{2,1})}{1-p_{1,2}-p_{2,1}}}
\end{eqnarray}
Subsequently, $p_x$ that satisfies this equality can be plugged into (\ref{eqninf}) and following some algebraic operations, the final capacity can be expressed as
\begin{eqnarray}
    C_2(p_{1,2},p_{2,1}) = \log_2\left(1+2^{\frac{h(p_{1,2})-h(p_{2,1})}{1-p_{1,2}-p_{2,1}}}\right) - \frac{(1-p_{2,1})h(p_{1,2}) + p_{1,2}h(p_{2,1})}{1-p_{1,2}-p_{2,1}} \label{M2capacity}
\end{eqnarray}
Finally, the ITR in bits/min can be found by $\frac{60}{T}C_2(p_{1,2},p_{2,1})$.

\subsubsection{\textcolor{black}{Extension to} General Case ``M symbols":}


\textcolor{black}{When there are more than two classes, the computations mentioned above become more complex as they require the calculation of partial derivatives and solving of transcendental equations. This complexity makes it impossible to obtain closed form results. In the past, there have been successful efforts to iteratively compute the capacity for discrete stationary channel models.}

For memoryless channels (independent choice of symbols) with finite input and output alphabets $\mathcal{X}$ and $\mathcal{Y}$ respectively, the capacity  can be computed by the Blahut-Arimoto (BA) algorithm \cite{arimoto1972,blahut1972}. On the other hand, in a typical speller task, due to the formation of language and words, the source will inherently have memory. The Blahut-Arimoto algorithm was also extended to channels with
memory and finite input alphabets and state spaces \cite{vontobel2004} such as Hidden Markov Models (HMM). However, modeling language with an HMM is quite challenging \cite{chiu2020} and can result in inordinate computation time and/or \textcolor{black}{only} an approximation for the capacity. 

\begin{figure}[!t]
\centerline{\includegraphics[width=0.95\columnwidth]{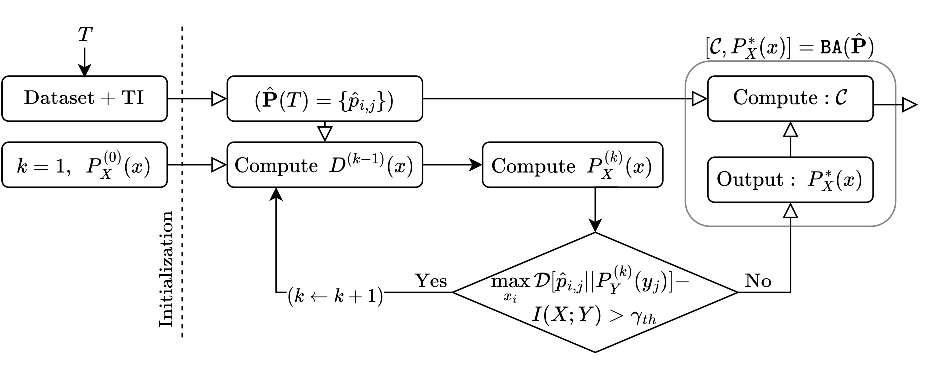}}
\caption{\textcolor{black}{For a given estimate of the channel transition statistics $\hat{\mathbf{P}}(T)$ based on a window length $T$, this figure illustrates the numerical computation of the capacity $\mathcal{C}$ that is used to estimate the information transfer rates. Note that this procedure also computes the optimal input distribution $P_X^*(x)$ that leads to $\mathcal{C}$. In BCI settings however, we typically need to optimize both the input and the channel transition statistics $\hat{\mathbf{P}}(T)$ simultaneously (see Discussion section).}}
\label{figBAlg}
\end{figure}

\textcolor{black}{We have employed a variant of} the BA algorithm, \textcolor{black}{which} is an iterative procedure \textcolor{black}{and} assumes an arbitrary input probability distribution function $P^{(0)}_X(x)$ in the beginning and optimizes it over multiple iterations \cite{Naja2009}. \textcolor{black}{Since access to the real channel transition statistics is unavailable, we used data to estimate transition probabilities as $\hat{p}_{i,j}$.} Let us express the non-normalized \textcolor{black}{estimated} input distribution for $x_i \in \mathcal{X}$ at the $(k-1)$-th step $(k \geq 1)$ as
\begin{eqnarray}
    D^{(k-1)}(x_i) &= P^{(k-1)}_X(x_i) \exp\left({\mathcal{D}\Big[\hat{p}_{i,j} || P^{(k-1)}_Y(y_j)\Big]}\right) \\
    &= P^{(k-1)}_X(x_i) \exp\left(\sum_{i=0}^N \hat{p}_{i,j} \log \left(\frac{\hat{p}_{i,j}}{P^{(k-1)}_Y(y_j)}\right)\right) \\
    &= P^{(k-1)}_X(x_i) \prod_{i=0}^N \left(\frac{\hat{p}_{i,j}}{P^{(k-1)}_Y(y_j)}\right)^{\hat{p}_{i,j}} \label{eqnD}
\end{eqnarray}
where $P^{(k-1)}_Y(y_j) = \sum_i P^{(k-1)}_X(x_i) \hat{p}_{i,j} $ and $\mathcal{D}[.||.]$ is the relative entropy (also known as Kullback-Leibler or KL divergence) \textcolor{black}{which is given for two probability distributions $P(x)$ and $Q(x)$ as  
\begin{eqnarray}
    \mathcal{D}[P || Q] = \sum_{x} P(x) \log{\frac{P(x)}{Q(x)}}.
\end{eqnarray}}
Then, the input distribution can be normalized and \textcolor{black}{finally} expressed as,
\begin{eqnarray}
    P^{(k)}_X(x_i) = \frac{D^{(k-1)}(x_i)}{\sum_j D^{(k-1)}(x_j)}. \label{eqnP}
\end{eqnarray}
Note that we can compute $D^{(k)}(x_i)$ and $P_X^{(k)}(x_i)$ iteratively using \textcolor{black}{equations (\ref{eqnD}) and (\ref{eqnP}) as given above}. \textcolor{black}{Ultimately}, iterations will cease if 
\begin{eqnarray}
    \max_{x_i} \mathcal{D}\Big[\hat{p}_{i,j} || P^{(k)}_Y(y_j) \Big] - \underbrace{\sum_{i} P_X^{(k)}(x_i) \mathcal{D}\Big[\hat{p}_{i,j} || P^{(k)}_Y(y_j) \Big]}_{I(X;Y)} < \gamma_{th}
\end{eqnarray}
holds for a given error threshold $\gamma_{th}$ (e.g., $\gamma_{th}=10^{-5}$). For the rest of our discussions, we refer to \textcolor{black}{this iterative process} as
 \begin{equation}
[\mathcal{C}, P^*_X(x)] = \texttt{BA}(\hat{\mathbf{P}}),
 \end{equation}
where \texttt{BA}$(.)$ refers to the algorithm \textcolor{black}{leveraging the estimated probabilities}, $\mathcal{C}$ is the capacity, \textcolor{black}{$\hat{\mathbf{P}} = \{\hat{p}_{i,j}\}$} and $P^*_X(x) \textcolor{black}{:= P_X^{\infty}(x)}$ is the optimal input distribution that maximizes the mutual information. \textcolor{black}{For better visualization, we have also provided the flow diagram of the iterative computation of the capacity in Fig. \ref{figBAlg}.}
\textcolor{black}{
\begin{remark}
This two-step process can be shown to converge through iterations of two different convex optimization problems and the convergence rate \textcolor{black}{is} shown to improve in a later study \cite{sinha2014}. 
\end{remark}
\begin{remark}
The expressions we have developed can be applied to channel models that have a number of outputs greater than $M$. For instance, Fig. \ref{fig1} depicts an example of a channel model with an ``erasure" output, which has $M+1$ possible outcomes.
\end{remark}}

\subsection{Fano's Inequality}
\textcolor{black}{Optimization of channel transition statistics may lead to impractical and unrealistic outcomes. To impose practical constraints, we invoke one of the well known bounds on the conditional entropy of the channel statistics, known as Fano's inequality. In essense,} Fano's inequality provides a lower bound for the probability of target identification error $\epsilon_M$ ($=1-\sum_i P_X(x_i) p_{i,i}$) due to information degradation via the channel induced by the BCI. The channel transition (conditional) probabilities $P_{Y|X}(y_j | x_i)$, emprically estimated by the confusion matrices, appear in the bound as follows,
\begin{eqnarray}
\epsilon_M \geq  \frac{H(Y|X) - h(\epsilon_M)}{\log_2(M-1)} \geq \frac{H(Y) - I(X;Y) - 1}{\log_2(M)} \label{FanosInequality}
\end{eqnarray}
where $h(\epsilon_M) = -\epsilon_M \log(\epsilon_M) - (1-\epsilon_M)\log(1-\epsilon_M)$ is the binary entropy function. Accordingly, for a fixed $\epsilon_M$, we can upper bound the conditional entropy as
\begin{eqnarray}
    H(Y|X) &\leq& \epsilon_M \log_2 \left(\frac{M-1}{\epsilon_M}\right) + (1-\epsilon_M) \log_2 \left(\frac{1}{1-\epsilon_M}\right) \label{FanosUnequality2} \\
    &=& h(\epsilon_M) + \epsilon_M \log_2(M-1)
\end{eqnarray}

Since in typical telecommunication applications the channel transition probabilities are given as part of the model, the symbol detection error is bounded. As can be seen, $\epsilon_M$ appears in both sides of the inequality (\ref{FanosInequality}). 
\textcolor{black}{
\begin{remark}
Note that the bound in (\ref{FanosInequality})  does not apply to $M=2$ case (zero denominator) and countably infinite sets. However later studies extended this bound to such corner cases \cite{Erdogmus2004, Ho2010}.  
\end{remark}
\begin{remark}
The bound on the conditional entropy given in (\ref{FanosUnequality2}) becomes $h(\epsilon_M)$ when $M=2$ and can be shown to be looser for larger $\epsilon_M$ (see our experimental results). 
\end{remark}}

\subsection{Target Identification}

It is conventional to divide TI methods \textcolor{black}{developped for SSVEP-based BCIs} into two broad overarching categories, namely \textit{supervised} and \textit{unsupervised} (training-free). Unsupervised techniques are particularly attractive since their use does not involve user-specific-calibration phase (long training cycles) and provides more versatility in everyday practices. On the other hand, supervised methods are shown to outperform the unsupervised early strategies such as Cannonical Correlation Analysis (CCA) \cite{Lin2006}. 

One of the most effective frequency recognition performance has been demonstrated by a number of spatial filtering techniques to isolate task-specific source activities from EEG signals. The task-related component analysis (TRCA) is one of prominent techniques proposed in literature which hypothesizes that there are distinct cortical sources in the brain which generates potentials upon the presentation of flickering stimuli. This idea originally applied to near-infrared spectroscopy (NIRS) \cite{Nakanishi2013}, \cite{Takana2014} and later proven useful for multivariate EEG data \cite{Nakanishi2018}.  Assuming $s(t) \in  \mathbb{R}$ to be the task-related, $n(t)$ to be the task-unrelated (noise and some other background brain activity) components, the multivariate EEG signal $\mathbf{x}(t) \in \mathbb{R}^{N_c}$ is formed as a result of a linear generative model as follows,
\begin{eqnarray}
    x_j(t) = a_j s(t) + b_j n(t) \textrm{ for } j=1,2,\dots,N_c
\end{eqnarray}
where $N_c$ is the number of channels. Spatial filtering is about extracting the task-related component $s(t)$ from a linear combinations of multiple channel output signals $\mathbf{x}(t)$, i.e., 
\begin{eqnarray}
    \tilde{s}(t) = \sum_j^{N_c} w_j x_j(t)  = \sum_j^{N_c} w_j a_j s(t) + w_j  b_j n(t)
\end{eqnarray}

Main idea behind TRCA is to optimize weight coefficients ($w_j$) so as to maximize inter-trial covariance (reproducibility) of time-locked biomedical data. If we denote the $h$-th trial of $y(t)$ as $y^{(h)}(t)$, then the sum of covariances of all possible combinations of trials can be expressed as
\begin{eqnarray}
    \sum_{h_1,h_2 = 1 \atop h_1 \not= h_2}^{N_t} \textrm{Cov} \left( y^{(h_1)}(t), y^{(h_2)}(t) \right) = \mathbf{w}^T \mathbf{S} \mathbf{w}
\end{eqnarray}
where $N_t$ is the total number of trials, $\mathrm{Cov(.)}$ is the covariance operator, $\mathbf{w} = (w_j)_{1\leq j \leq N_c}$ represents the weights of the spatial filter and $\mathbf{S} = \left(S_{j_1,j_2}\right)_{1 \leq j_1,j_2 \leq N_c}$ is given by
\begin{eqnarray}
    S_{j_1,j_2} = \sum_{h_1,h_2 = 1 \atop h_1 \not= h_2}^{N_t} \textrm{Cov} \left( x_{j_1}^{(h_1)}(t), x_{j_2}^{(h_2)}(t) \right)
\end{eqnarray}
For a finite solution, TRCA maximizes $\mathbf{w}^T \mathbf{S} \mathbf{w}$ subject to variance constraint, i.e., Var($y(t)$) $ =  \mathbf{w}^T \mathbf{Q} \mathbf{w} \leq 1$. The solution is given by the following unconstrained optimization problem
\begin{eqnarray}
    \mathbf{w}^* = \arg \max_{\mathbf{w}} \frac{\mathbf{w}^T \mathbf{S} \mathbf{w}}{\mathbf{w}^T \mathbf{Q} \mathbf{w}} 
\end{eqnarray}
which can be recognized as a generalized Eigenvalue problem. 

By creating a spatial mapping that projects the multivariate EEG data onto a standard SSVEP representation space, the sum of squared correlations (SSCOR) framework \cite{Kumar2019} seeks to identify a session independent representation of SSVEP response. In this case, we express the optimization problem as 
\begin{eqnarray}
    \mathbf{w}^*_\mathcal{X},(\mathbf{w}_i^*) = \arg \max_{\mathbf{w}_\mathcal{X}, \mathbf{w}_i} \sum_{i=1}^{N_t}\left[ \mathbf{w}_\mathcal{X}^T \textrm{Cov} \left( x^{(\mathcal{X})}(t), x^{(i)}(t) \right) \mathbf{w}_i \right]^2
\end{eqnarray}
where 
\begin{eqnarray}
    x^{(\mathcal{X})}(t)= \frac{1}{N_t} \sum_{i=1}^{N_t} x^{(i)}(t)
\end{eqnarray}
is the template signal calculated for each target frequency, separately. Again for the sake of obtaining a finite solution and put the optimization problem into a generalized eigenvalue framework, we use the set of constraints for $\forall i$, $\mathbf{w}_i^T \textrm{Cov} \left( x^{(i)}(t), x^{(i)}(t) \right) \mathbf{w}_i = 1$.  

Note that there could be other spatial filtering techniques that can generate the filter weights ($\mathbf{w}$) as an application of generalized Eigenvalue problem \cite{Wong2020}.  However, the arguments for the detection logic is common to all.  Thus, in our work we will focus on these two TIs when we report our results. 

After determining the spatial filter weights, for a given single-trial test sample $\mathbf{X}$, the classification decision is made in favor of the frequency $f_n \in \{f_1,\dots,f_{N_f}\}$ based on the  Pearson's correlation coefficient ($\rho$) as a solution to
\begin{eqnarray}
    \tau = \arg \max_n \rho \left( \mathbf{X}^T \mathbf{w}_n, \mathcal{X}_n^T \mathbf{w}_n \right), \label{decision_ssvep}
\end{eqnarray}
where $N_f$ is the total number and $n$ is the index of the stimulation frequency. 
\textcolor{black}{Assuming the $i$-th maximum correlation for the frequencies ${f_n}$ is denoted by $\tau_i$, consider a given correlation threshold $\tau_t$. If there exist multiple frequencies that satisfy the condition $\rho \left( \mathbf{X}^T \mathbf{w}_n, \mathcal{X}_n^T \mathbf{w}_n \right) \leq \tau_t$, i.e., $\forall i$, $\tau_i \leq \tau_t$, then it might be advisable to refrain from making any definitive decisions. Instead, the threshold value $\tau$ could be replaced with $e$, which represents an ``erasure" (see Fig. \ref{fig1}).} Finally, if we concatenate all weights to construct the matrix $\mathbf{W} = [\mathbf{w}_1 \ \mathbf{w}_2 \ \dots \ \mathbf{w}_{N_f}]$ and replace it with $\mathbf{w}_n$ in Equation (\ref{decision_ssvep}), we obtain an ensemble TI algorithm. Same idea can be applied to both TI methods. 

\section{Experimental Results}

\subsection{Datasets} 

We have used two well-known datasets in the literature, namely, Benchmark \cite{Wang2016} and Beta datasets \cite{liu2020} to analyze/compare two known target identification algorithms, namely the ensemble extensions of TRCA and SSCOR, based on the conventional as well as proposed ITR definitions. 

Benchmark dataset was collected from 35 subjects with normal/corrected-to-normal vision on a 40-character (26
English alphabet letters, 10 digits, and 4 other symbols) speller task using a 64-channel EEG recorder. Each subject was shown target characters in distinct trials, where characters flicker at frequencies $8$-$15.8$ Hz with $0.2$ Hz increments and phases $0$-$1.5\pi$ with $0.5\pi$ increments, where both increments are proportional to stimulus index. Each trial began with a visual cue that was shown on the screen for $0.5$ seconds to direct the subject's gaze to the intended target, followed by $5$ seconds of stimulation and a final trailing $0.5$-second offset, respectively. The observation window includes gaze length as well as the signal length after the stimuli onset. At the preprocessing stage, the recorded signals were downsampled to $250$Hz. We have assumed $130$ ms visual pathway delay for both datasets. In this study, the following set of 9 channels
(OZ, O1, O2, PZ, POZ, PO3 PO4, PO5 and PO6) are considered since they are reported to be most reflective of neural activity due to tasks performed in the experiment \cite{Wang2016}. In Beta dataset on the other hand, $70$ subjects participated in four blocks of a cued-spelling task on a QWERTY virtual keyboard. The stimulation duration is $2$ seconds for the first $15$ subjects whereas the rest of the other subjects are stimulated for $3$ seconds. Sixty-four channels of EEG data were collected by SynAmps2 (Neuroscan Inc.) data acquisition/amplifier system at a sampling rate of $1000$Hz, which were later downsampled to $250$Hz. The rest of the settings are the same (channels used, electrode locations, gaze-shift times, visual pathway latency, etc.). Since the trials were carried out outside a laboratory setting, the Signal-to-Noise Ratio (SNR) of EEG in Beta dataset is measured to be lower than that of the Benchmark dataset. \textcolor{black}{To obtain additional details about the pre-processing procedures, we suggest that the readers consult the publications in which the creators of these datasets initially introduced them.}

\subsection{Filter Banks}
In our work, we have leveraged filter banks to decompose the  EEG signals into five overlapping sub-bands to make use of the independent information found in the harmonics. A target detection technique is used independently for each of the sub-bands. The cut-off frequency range for the sub-bands based on the EEG data bandwidth is set between $b \times 8$ Hz and $90$ Hz, where $b \in \{1,2,3,4,5\}$ \cite{Chen2015}. As the Target Identification (TI) algorithm,  we have employed two competing approaches, namely the TRCA \cite{Nakanishi2018} and SSCOR  \cite{Kumar2019} due to their high performance relative to other methods with reasonable complexity (in terms of the number of parameters to optimize).

As mentioned before, the design methodology behind TRCA was the hypothesis that the single-trial EEG data can be reconstructed as a linear sum of multiple time series from
different cortical sources \cite{Onton2006}. Therefore, TRCA is used as a technique for highlighting the task-related elements embedded in the EEG signals through enhancing repeatability among different time-locked activities across trials. The covariance between different trials is maximized to ensure such repeatability.  On the other hand, SSCOR transforms SSVEP signals to a common representation space through the optimization of the individual SSVEP templates. Similar to TRCA, SSCOR also achieves space transformation.  In both of these methods, we learn a spatial filter for each frequency (character) \cite{Wong2020}. In our work, we have also employed an ensemble technique for both methods where all spatial filters belonging to different frequencies are concatenated for a performance boost. To determine a final score (a single correlation coefficient) for classification, the correlation coefficients of the sub-band components are combined using the weighted sum of squares approach \cite{Chen2015}.

\begin{figure}[!t]
\centerline{\includegraphics[width=0.9\columnwidth]{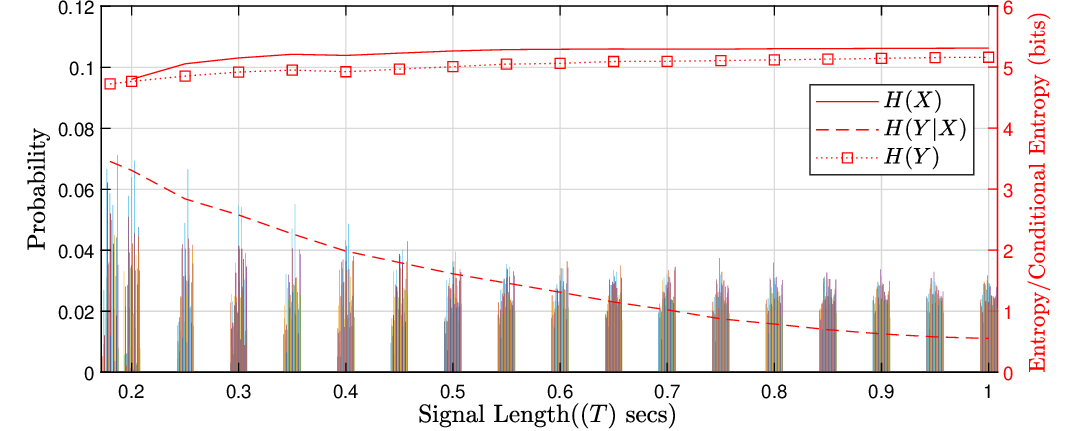}}
\caption{Capacity achieving distributions for different signal lengths (sec) as well as their entropies for TRCA method using Benchmark dataset. \textcolor{black}{Each bar plot consists of 40 bars corresponding to the the probabilities assigned to each input symbol in the speller application.}}
\label{fig11}
\end{figure}

\subsection{Evaluation Process} 
For each subject, we evaluated performance in a leave-one-trial-out fashion as in 
the past literature i.e., the TI algorithms are trained over $u-1$ trials and tested on the remaining trial for all $u$ different combinations for $u=6$ in Benchmark and $u=4$ in Beta datasets. When we present average subject performance, test performances are combined (classification outcomes) in a single confusion matrix, which is normalized to obtain the estimate of the channel transition probability matrix $\tilde{\mathbf{P}}$ with $M=40$. In some of the past works c.f. \cite{Nakanishi2018}, \cite{Kumar2019}, using conventional definition, ITR is calculated for each subject and the average of these values (with the standard error) is reported despite the nonlinear dependency of the definition on the accuracy. This type of averaging typically leads to larger ITR values. In our work, we report all ITR results after averaging the accuracy values (similarly false positive and negatives etc.) over all subjects before calculating and reporting the final ITR. This way, we also ensure that the aggregate human \textcolor{black}{accuracy} performance is translated into a \textcolor{black}{single} ITR metric. Note that with the proposed definition, it would be unrealistic to expect the system to optimize the input distributions for each subject separately in order to attain higher ITRs. To address this point however, we have also demonstrated via violin plots by calculating the individual ITRs using both definitions \textcolor{black}{(conventional v.s. proposed)} for all observation windows. \textcolor{black}{To this end}, we use individual data to estimate the channel transition statistics for each subject and observation window, separately. 
We have leveraged \texttt{BA}(.) algorithm to compute the capacity--\textcolor{black}{achieving} input distributions and the corresponding capacity value for signal lengths of $0.18,0.2,0.25,0.3,\dots,1$ seconds, whereby the conditional entropy of  $\tilde{\mathbf{P}}$ \textcolor{black}{is observed to  comply} with the Fano's inequality. 

\begin{figure}[!t]
\centerline{\includegraphics[width=0.9\columnwidth]{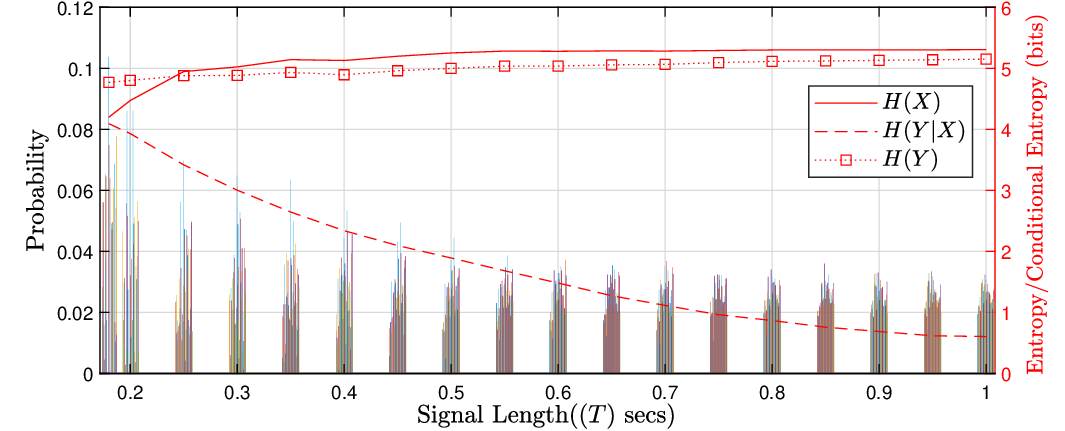}}
\caption{Capacity achieving distributions for different signal lengths (sec) as well as their entropies for SSCOR method using Benchmark dataset. \textcolor{black}{Each bar plot consists of 40 bars corresponding to the the probabilities assigned to each input symbol in the speller application.}}
\label{fig12}
\end{figure}

\subsection{Numerical Results} 

\subsubsection{\textcolor{black}{ITR} performance averaged over subjects:} In Figs. \ref{fig11} and \ref{fig12}, using Benchmark dataset, we demonstrate capacity--achieving distributions (with channel transition statistics \textcolor{black}{estimated using the data of all} subjects) as bar plots for each signal length. \textcolor{black}{Also included in the same figures are} the corresponding input entropy $H(X)$, output entropy $H(Y)$ and the conditional entropy $H(Y|X)$ that characterizes the channel transition statistics for both TI methods, respectively. 
\textcolor{black}{It is apparent that, as the observation window length expands, there is an increase in both input entropy $H(X)$ and output entropy $H(Y)$. However, the channel transition probabilities begin to display more intricate patterns, such as decreased symmetry and a more evenly distributed set of probabilities (the far right bar plot), resulting in lower conditional entropy.} As a result, the \textcolor{black}{mutual information} of the induced channel (\textcolor{black}{i.e.,} $H(Y)-H(Y|X)$) increases with growing signal length in both techniques. As can be seen, particularly at low signal lengths, SSCOR fails to enhance the channel capacity due to poorer reduction in the conditional entropy $H(Y|X)$. However, as the observation window \textcolor{black}{length} increases, the reduction in  $H(Y|X)$ for both TIs become on par, almost equating the capacity gain at around an observation window of one seconds. We observed similar trends using the Beta dataset. 

\begin{figure}%
    \centerline{
    \subfloat{{\includegraphics[width=7.9cm]{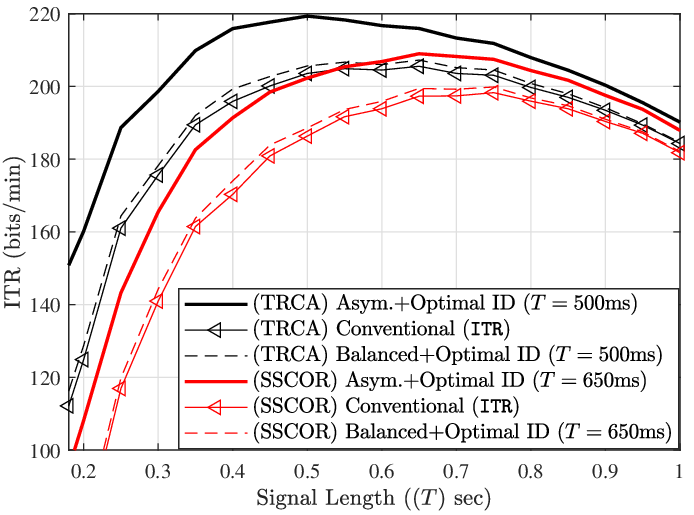} }}%
    \subfloat{{\includegraphics[width=7.9cm]{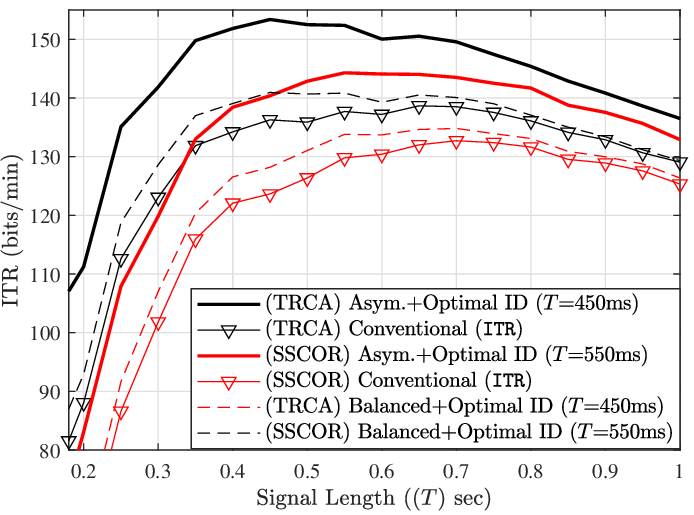} }}
    }%
    \caption{Average ITR as a function of signal Length (sec) for different TIs for Benchmark dataset (left) and Beta dataset (right). The plots clearly show the difference between the conventional definition as well as the proposed definitions. Since input distribution depends on the window length $T$, we have also expressed which time window the ID is optimal for.}
    \label{fig2}
\end{figure}

We have also provided the ITR results using the \textcolor{black}{proposed scheme as well as the}  confusion matrices in Fig. \ref{fig2} using both datasets. Since the ITR is maximized at the signal length of $\approx 0.5$ secs for TRCA and $\approx 0.65$ secs for SSCOR for Benchmark dataset (and $\approx 0.45$ secs for TRCA and $\approx 0.55$ secs for SSCOR for Beta dataset), we use the \textcolor{black}{capacity-achieving or} optimal input distribution (ID) of the signal lengths $0.5$ ($0.45$) and $0.65$ ($0.55$) secs, respectively, for all \textcolor{black}{window} lengths to obtain the ITR for the TI algorithms (\textit{Asym.}+\textit{Optimal ID}). For comparison purposes, we have also included the conventional ITR definition in Fig. \ref{fig2}, which neither takes the input distribution nor the asymmetry the channel introduces into account and consequently underestimates the maximum information transfer rate that can be achieved over the induced channel. To investigate it statistically, we have carried out paired $t$-test and $f$-test (two-tailed) to determine whether the proposed ITR definition is \textcolor{black}{statistically} different from the conventional. \textcolor{black}{Recall that the $f$-test and $t$-test are statistical tests that are used to determine whether there is a significant difference between the means and variances (respectively) of two groups or samples.} Using Benchmark dataset, both methods showed dramatic mean differences ($p \approx 3.97\times 10^{-8}$ for SSCOR and $p \approx 3.7\times 10^{-7}$ for TRCA). However, there were no significant variational differences between the different ITR definitions ($F(17,17)=1.25, p>0.05$ for SSCOR and $F(17,17)=1.47, p>0.05$ for TRCA). This clearly demonstrates that although the trend is almost the same for both TI algorithms, the actual ITR experience is meaningfully different than the reported average ITR results in the literature. 

\subsubsection{\textcolor{black}{Channel asymmetry and ITR performance}:}  To be able to demonstrate the effect of asymmetry of the channel on the ITR, we have used the optimal ID while keeping the accuracy intact and divide the error probabilities equally over all other non-target classes for each class (character) i.e., $p_{i,j}=\frac{1-p}{M-1}$ for all $i,j$ satisfying $i\not=j$. We have \textcolor{black}{dubbed} this scheme ``\textit{Balanced} transition probability matrix" and used it with the optimal ID for all signal lengths (\textit{Balanced}+\textit{Optimal ID}). \textcolor{black}{Upon analyzing the results, it becomes evident that the effect of the asymmetry in probability transition characteristics on the final ITR outcomes is much more pronounced in comparison to the impact of the non-uniformity of the input distribution.}

\begin{figure}%
    \centerline{
    \subfloat{{\includegraphics[width=7.9cm]{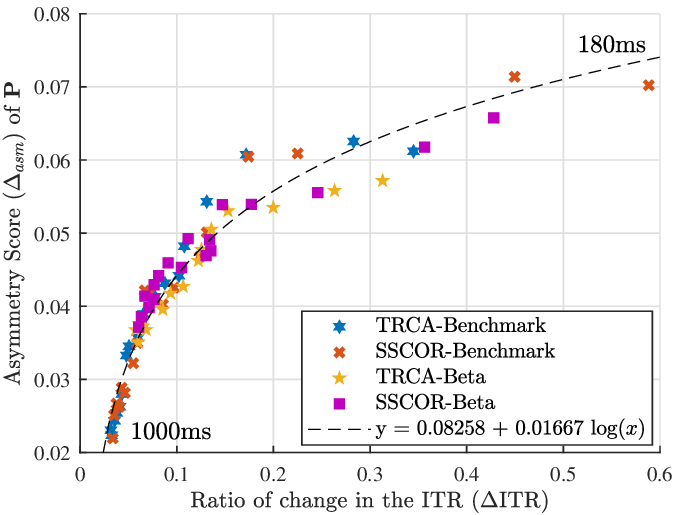} }}%
    \subfloat{{\includegraphics[width=7.9cm]{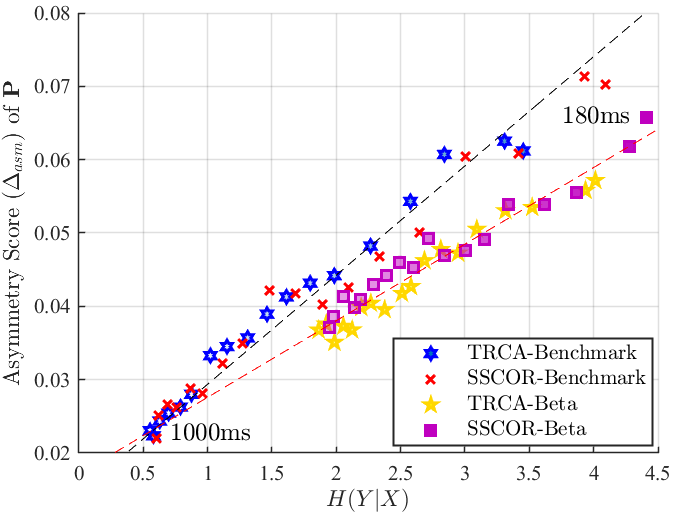} }}%
    }
    \caption{On the left, we show asymmetry score $\Delta_{asm}$ as a function of the ratio of change in the ITR with the new definition for all $T$. On the right, we have also depicted the asymmetry score $\Delta_{asm}$ as a function of the conditional entropy of the induced channel. We have varied the transparency of the markings from dark to light as we change the observation time window from 0.18 seconds to 1 seconds.}
    \label{figITR2}
\end{figure}

On the other hand, to quantify the degree of asymmetry, we have used the largest singular value of the skewed Laplacian of the channel transition matrix, namely $\Delta_{asm} = (\Gamma - \Gamma^T)/2$ \cite{li2012}, where  $\Gamma = \Phi^{1/2} (\mathbf{I} - \mathbf{P}) \Phi^{-1/2}$, $\mathbf{I}$ is the identity matrix,
$$\Phi^{1/2} = \left[ \matrix{  \sqrt{\phi_1} & 0 & \dots  & 0 \cr
0 & \sqrt{\phi_2} & \dots  & 0 \cr
\vdots & \vdots & \vdots & \ddots \cr
0 & 0 & \dots  & \sqrt{\phi_M} \cr} \right],$$
and $[\phi]_{1 \leq i \leq M}$ are the stationary probabilities\footnote{\textcolor{black}{In our work, we use the estimates of these probabilities, computed based on a given dataset.}} of a Markovian process whose transitions are governed by the channel transition statistics. In Fig. \ref{figITR2} (left), we depicted the degree of asymmetry $\Delta_{asm}$ as a function of the ratio of change in the ITR ($\Delta$ITR) using the proposed definition. More specifically, 
\begin{eqnarray}
    \Delta \mathrm{ITR} 
\triangleq \frac{\textrm{(Asym.+Optimal ID)} - \textrm{(Conventional)}}{\textrm{(Conventional)}}. 
\end{eqnarray}


As can be clearly seen from Fig. \ref{figITR2} (left), asymmetry increases the ITR gain, and it turns out to satisfy a logarithmic relationship. The parameters of this estimate is explicitly given in the same figure using regression. This suggests that although the asymmetry in the channel transition statistics helps with increasing the rate of communication, it also quickly saturates due to increased conditional entropy of the channel and reduced accuracy. Note that the conditional entropy is bounded above by the Fano's inequality as expressed in (\ref{FanosUnequality2}) which forms an upper bound on $\Delta_{asm}$. One of the interesting conclusions is that particularly at short window lengths, stimuli design that generates more asymmetric channel transition characteristics is likely to improve $\Delta$ITR, higher improvement ratio with respect to the conventional definition. However, the relationship between $\Delta_{asm}$ and $H(Y|X)$ clearly suggests that as we have larger observation windows, to be able to maximize the mutual information, system prefers to have less asymmetry in the channel transition characteristics (to minimize the conditional entropy) and maximize the input entropy via uniform distribution. 

\begin{figure} [t!]
    \subfloat{{\includegraphics[width=16cm]{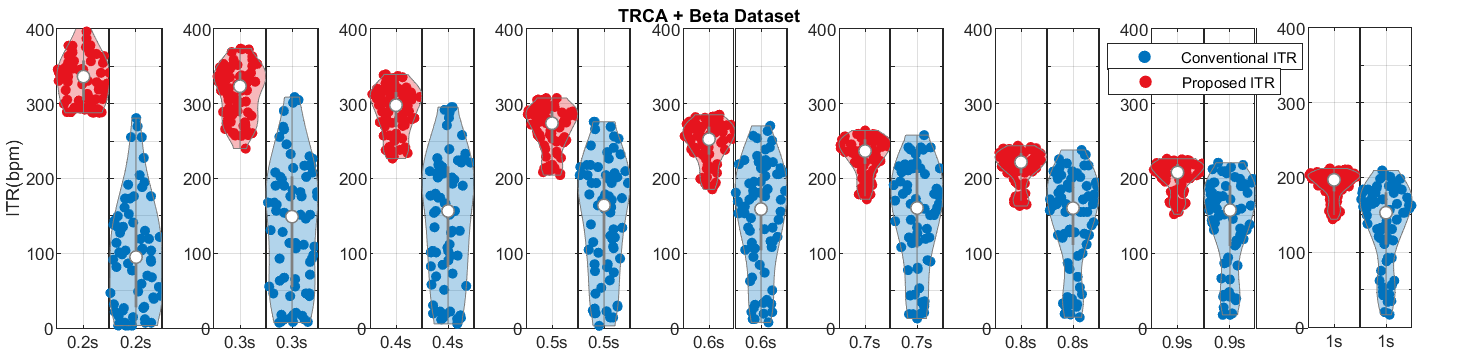} }}%
    \\
    \subfloat{{\includegraphics[width=16cm,height=4.4cm]{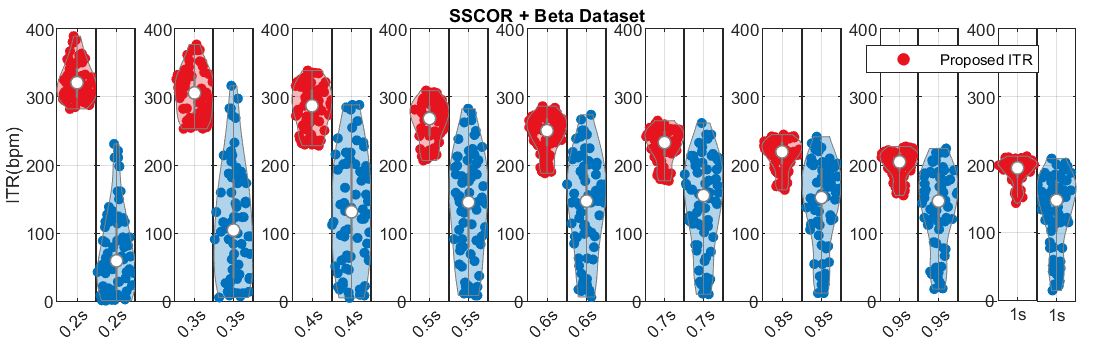} }}%
    \caption{Proposed ITR v.s. conventional ITR for subject-specific input customization using two TIs and the Beta Dataset. Input customization can be shown to be effective even with longer observation windows.}
    \label{figviolin}
\end{figure}

On the other hand in Fig. 5 (right), we have illustrated $\Delta_{asm}$ as a function of conditional entropy $H(Y|X)$. As anticipated, there is a monotonic relationship (presented with a linear regression) in between and both the degree of asymmetry and conditional entropy increase as the observation window length grows. One of the interesting observations is that the dataset has more significant effect on the slope of the relationship compared to target identification algorithms. Different slopes can also be interpreted as an indicator of the difficulty of the datasets (TIs perform worse with Beta dataset compared to Benchmark) and reduced SNR while recording these EEG datasets.

\begin{figure}%
    \centering
    \subfloat{{\includegraphics[width=16cm]{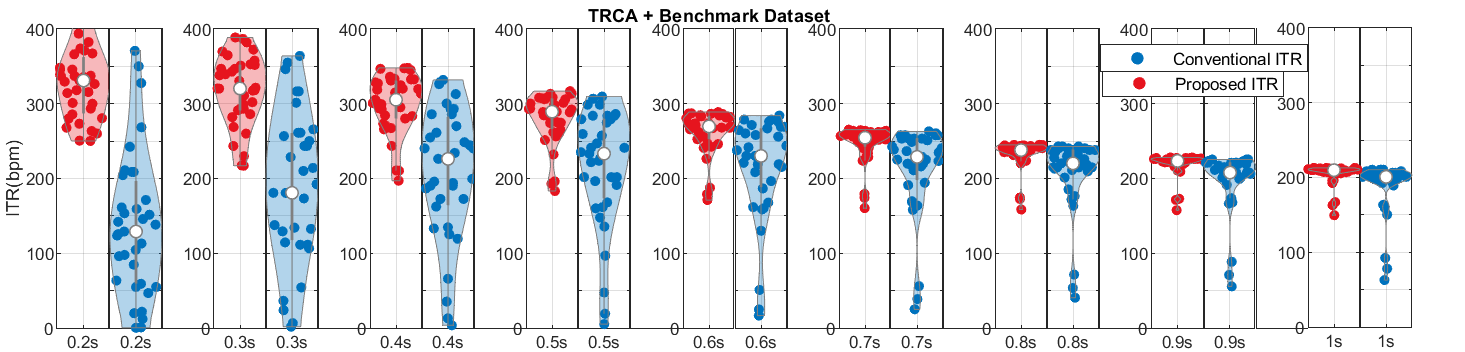} }}%
    \\
    \subfloat{{\includegraphics[width=16cm]{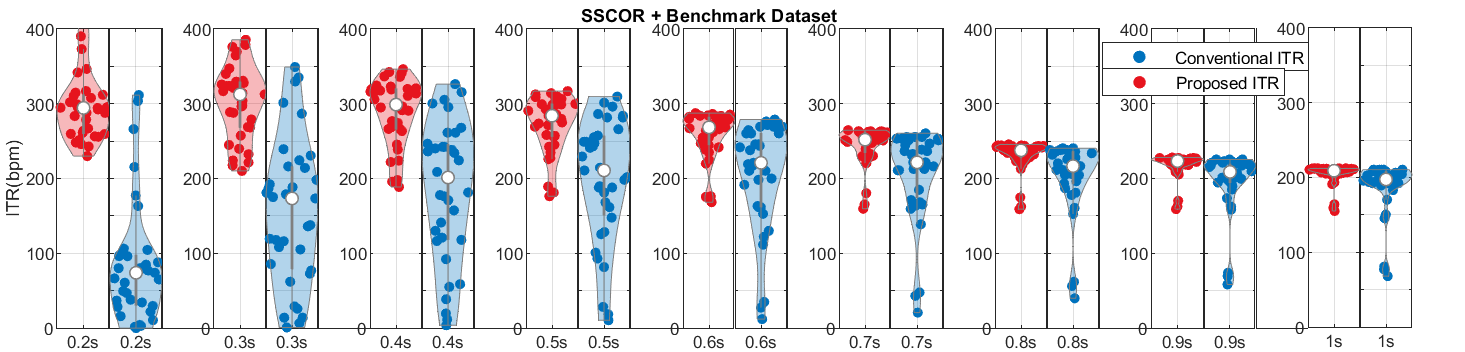} }}%
    \caption{Proposed ITR v.s. conventional ITR for subject-specific input customization using two TIs and the Benchmark Dataset. Input customization can be shown to be effective even with longer observation windows.}
    \label{figviolin2}
\end{figure}

\subsubsection{\textcolor{black}{ITR} performance customized to each subject:} 
\begin{figure}%
\centerline{
\subfloat{{\includegraphics[width=19.2cm]{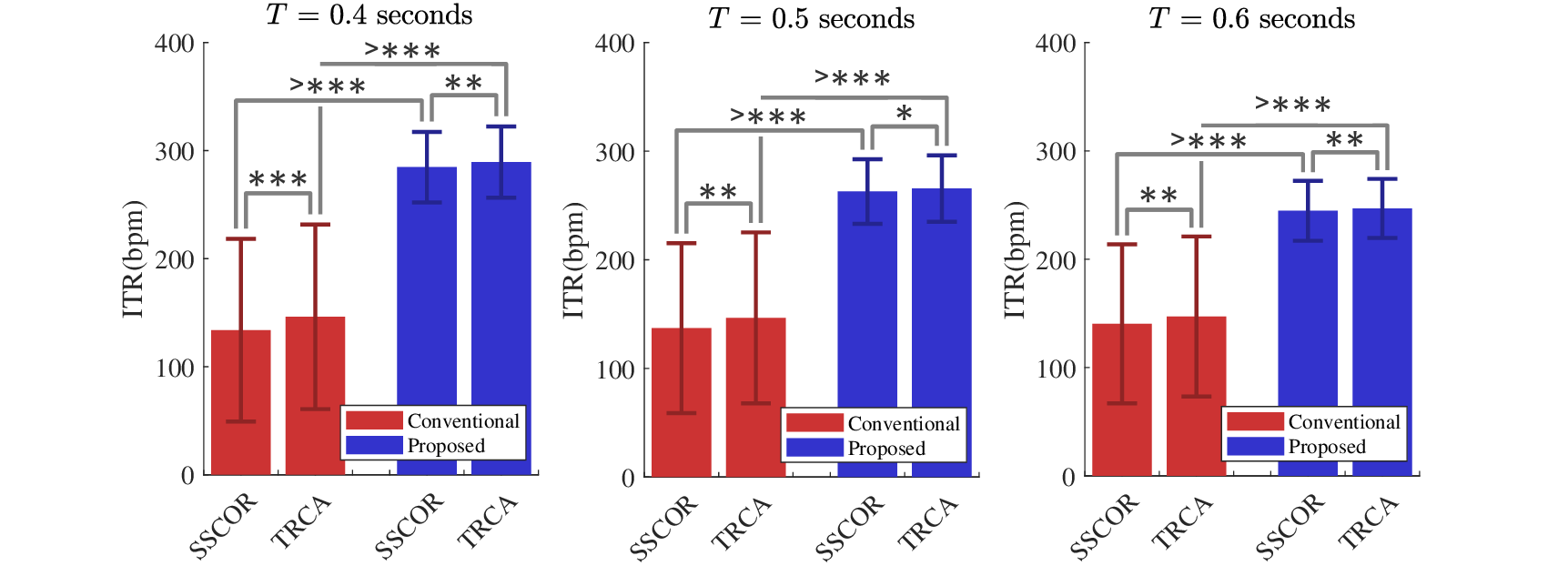} }}%
}
\caption{\textcolor{black}{Performance (ITR) comparisons using different definitions (Conventional v.s. Proposed) using three window lengths (that provides the best average ITR for the Beta dataset) along with the results of the statistical tests.}}
\label{figTestPlot}
\end{figure}

We have also investigated via violin plots in Figs. \ref{figviolin} and \ref{figviolin2} (considering all 70 subjects in the Beta,and all 35 subjects in the Benchmark individually as data points, respectively) whether the proposed ITR computed per subject (estimating the channel transition statistics based on individual subject data and finding the optimal input distribution for each subject) differs significantly compared to conventional definition for the two different TIs using both datasets. There are two important observations about the performance of TIs as a function of observation window ($0.2,0.3,\dots,1$ seconds). First, customizing the input distribution can tremendously help with the experienced ITR as compared to conventional definition. Particularly, at shorter observation windows, the difference is quite significant  for both TIs tested over both datasets.  Second observation is that even though the two TIs seem to differ in performance, more so with shorter observation windows, using the conventional definition, their ITR performance do not show significant difference using the proposed ITR definition with input distribution optimized across all observation windows. To show that statistically, using Beta dataset, we have conducted right-tailed paired $t$-test with the alternative hypothesis being the mean of TRCA performance is greater than that of the SSCOR performance. \textcolor{black}{Using} $95\%$ confidence interval, we have presented in Table 1 the corresponding critical as well as the associated $p$ values. As can be seen, lower critical value and higher $p$ values for the proposed ITR difference indicates the minor variation in the performance of TRCA and SSCOR algorithms. In addition, we have also conducted $f$-test to measure the variational difference of both algorithm's performances. As can be seen in Table 1, our $f$-statistic ($F(69,69)$) demonstrates (for all $T$ and $p \geq 0.041$) that the variational differences are not significant. \textcolor{black}{To corroborate our visual aids and put our results into perspective, we have additionally charted the average ITR for both conventional and proposed definitions, as a function of window lengths $(T \in {0.4,0.5,0.6})$ in Fig. \ref{figTestPlot}, accompanied by their corresponding $p$-value thresholds of $*:10^{-2}, **:10^{-4}, ***:10^{-6}$. The outcomes of the experiment demonstrates that the proposed definition mitigates the performance variability across diverse subjects. Moreover, it attenuates the significance of the mean performance disparity between the two TIs, namely SSCOR and TRCA.}

\begin{table}[] 
\centering
\footnotesize{
\begin{tabular}{l|lll|lll|}
\cline{2-7}
& \multicolumn{3}{l|}{\cellcolor[HTML]{C0C0C0}\begin{tabular}[c]{@{}l@{}}Proposed ITR Differences \end{tabular}} & \multicolumn{3}{l|}{\cellcolor[HTML]{C0C0C0}\begin{tabular}[c]{@{}l@{}}Conventional ITR Differences \end{tabular}} \\ \hline
\multicolumn{1}{|l|}{\cellcolor[HTML]{C0C0C0}\begin{tabular}[c]{@{}l@{}}Window \\ Length\\ $T$ (sec)\end{tabular}} & \multicolumn{1}{l|}{$p$ value}       & \multicolumn{1}{l|}{\begin{tabular}[c]{@{}l@{}}$t$-test\\ CV\end{tabular}}     & $F(69,69)$     & \multicolumn{1}{l|}{$p$ value}      & \multicolumn{1}{l|}{\begin{tabular}[c]{@{}l@{}}$t$-test\\CV\end{tabular}}     & $F(69,69)$     \\ \hline
\multicolumn{1}{|l|}{\cellcolor[HTML]{C0C0C0}0.2}                                                              & \multicolumn{1}{l|}{2.87e-11}      & \multicolumn{1}{l|}{8.1}                                                     & 0.83         & \multicolumn{1}{l|}{1.49e-17}     & \multicolumn{1}{l|}{24.41}                                                        & 0.865        \\ \hline
\multicolumn{1}{|l|}{\cellcolor[HTML]{C0C0C0}0.3}                                                              & \multicolumn{1}{l|}{1.45e-9}       & \multicolumn{1}{l|}{6.61}                                                    & 0.59         & \multicolumn{1}{l|}{2.84e-11}     & \multicolumn{1}{l|}{17}                                                           & 0.614        \\ \hline
\multicolumn{1}{|l|}{\cellcolor[HTML]{C0C0C0}0.4}                                                              & \multicolumn{1}{l|}{3.68e-6}       & \multicolumn{1}{l|}{3.08}                                                    & 0.577        & \multicolumn{1}{l|}{3.48e-07}     & \multicolumn{1}{l|}{8.61}                                                         & 0.58         \\ \hline
\multicolumn{1}{|l|}{\cellcolor[HTML]{C0C0C0}0.5}                                                              & \multicolumn{1}{l|}{1.49e-3}       & \multicolumn{1}{l|}{1.24}                                                    & 0.602       & \multicolumn{1}{l|}{5.6e-06}      & \multicolumn{1}{l|}{6.14}                                                         & 0.57         \\ \hline
\multicolumn{1}{|l|}{\cellcolor[HTML]{C0C0C0}0.6}                                                              & \multicolumn{1}{l|}{6.21e-5}       & \multicolumn{1}{l|}{1.31}                                                    & 0.552        & \multicolumn{1}{l|}{1.98e-05}     & \multicolumn{1}{l|}{4.23}                                                         & 0.57         \\ \hline
\multicolumn{1}{|l|}{\cellcolor[HTML]{C0C0C0}0.7}                                                              & \multicolumn{1}{l|}{9.2e-05}       & \multicolumn{1}{l|}{1.13}                                                    & 0.566        & \multicolumn{1}{l|}{1.92e-05}     & \multicolumn{1}{l|}{3.59}                                                         & 0.568        \\ \hline
\multicolumn{1}{|l|}{\cellcolor[HTML]{C0C0C0}0.8}                                                              & \multicolumn{1}{l|}{3.4e-3}        & \multicolumn{1}{l|}{0.52}                                                    & 0.579        & \multicolumn{1}{l|}{7.38e-4}      & \multicolumn{1}{l|}{2.13}                                                         & 0.559        \\ \hline
\multicolumn{1}{|l|}{\cellcolor[HTML]{C0C0C0}0.9}                                                              & \multicolumn{1}{l|}{2.51e-3}       & \multicolumn{1}{l|}{0.55}                                                    & 0.588        & \multicolumn{1}{l|}{1.43e-3}      & \multicolumn{1}{l|}{1.903}                                                        & 0.573        \\ \hline
\multicolumn{1}{|l|}{\cellcolor[HTML]{C0C0C0}1.0}                                                              & \multicolumn{1}{l|}{8.04e-3}       & \multicolumn{1}{l|}{0.32}                                                    & 0.572        & \multicolumn{1}{l|}{7.36e-4}      & \multicolumn{1}{l|}{1.905}                                                        & 0.567        \\ \hline
\end{tabular}
}
\label{table1}
\caption{Critical and $p$ values for the right-tailed paired $t$-tests ($95\%$ confidence) for 70 subjects of Beta data set using both definitions of ITR. The table also shows the $f$-statistic for $p \geq 0.041$ for the same confidence interval. CV: Critical Value.} 
\end{table}

\section{Discussion}

In this work, we have investigated the conventional definition of ITR, highlighted the deficiencies and proposed to use an algorithmic approach for more accurate computation of the information transfer. One conclusion we derive from this study is that the effect of the input distribution changes on the experienced ITR is quite minor when the performance is averaged over all subjects. This implies that in an online speller task, for instance, we typically would not have the option to change the ID and yet the TI algorithm would be operating close to the capacity--maximum ITR. However, the asymmetry (proportion of false positives and negatives) in the channel significantly changes the ITR in a typical BCI-based communication setting. Therefore, target identification algorithms, while in the design phase, should not solely optimize the accuracy performance. Note that ``accuracy" and ITR have long been used interchangeably in the SSVEP-based BCI literature and it is the only \textcolor{black}{objective} performance indicator of a TI that  appears in the conventional ITR definition (see Equation (\ref{eqn1ITR})). 

\begin{algorithm}[b!] 
\caption{Joint calculation of $\textbf{P}^* = \{p_{i,j}^*\}$ and $P^{*}_X(x)$ in an iterative manner.} 
\label{alg1} 
\begin{algorithmic}[1] 
    \REQUIRE $\sum_j p_{i,j} = 1, 0 \leq  p_{i,j} \leq 1, 0 \leq P_X(x_i) \leq 1$. 
    \ENSURE $\epsilon_M = 1 - \sum_{x_i} P_X(x_i) p_{i,i}$.
    \STATE $P_X(x_i) \Leftarrow \frac{1}{M}$ for all $i \in \{1,2,\dots,M\}$. 
    \STATE $p_{i,j} \Leftarrow \texttt{RAND(0,1)}$
    \STATE $p_{i,j} \Leftarrow p_{i,j}/\sum_j p_{i,j}$ \COMMENT{$\triangleright$ Normalization}
    \WHILE{$\sum_{x_i} P_X(x_i)p_{i,i} \leq 1-\epsilon_M$ and $H(Y|X) \leq h(\epsilon_M) + \epsilon_M \log_2(M-1)$} 
        \STATE $\hat{\textbf{P}} = arg \max_{{\atop \textbf{P}}} I(X,Y)$
        \COMMENT{$\triangleright$ Fix $P_X(x)$ and optimize $\textbf{P}$}
       \STATE $[\mathcal{C}_{DMC}, P_X(x)] =$ $\texttt{BA}(\hat{\textbf{P}})$ \COMMENT{$\triangleright$ Fix $\textbf{P}$, optimize $P_X(x)$}
    \ENDWHILE
    \STATE $\textbf{P}^* \Leftarrow \hat{\textbf{P}}$, $P^{*}_X(x) \Leftarrow P_X(x)$ 
\end{algorithmic}
\end{algorithm}

On the other hand, our results lead us to consider the possibility of alternative (better) channel transition (confusion) matrices. In other words, given an average target accuracy level ($1-\epsilon_M$) and a fixed ID ($P_x(x)$), one can optimize the channel transition probability matrix such that the channel mutual information is maximized i.e. to achieve the maximum ITR subject to conditional entropy bound of the channel statistics, given by the Fano's inequality. With this study, we can show the potential of this approach for the binary classification scenario as follows.

Let us consider a special case, a binary character transmission i.e. $M=2$. First, from (\ref{M2capacity}), we observe the symmetry $C_2(p_{1,2},p_{2,1}) = C_2(p_{2,1},p_{1,2})$ which is minimized for a given average 
accuracy target $A<1$ (or a classification error $\epsilon_M >0$) when $p_{1,2} = p_{2,1} = 1-A$. On the other hand, capacity is maximized when $(p_{1,2},p_{2,1}) \approxeq  (2(1-A),0)$ or  $(p_{2,1},p_{1,2}) \approxeq  (0,2(1-A))$ with equality if the input distribution is uniform i.e., $P_X(x) = \frac{1}{M}$. In other words, the ITR is maximized when either Precision or Recall is unity which can be obtained by playing with the parameters of the TI algorithm i.e., replacing the separating hyperplanes of the classifiers. However this argument does not take into account the natural bounds on the classification error and asymmetry of the channel statistics. 

For a given classification \textcolor{black}{target} error \textcolor{black}{rate} $\epsilon_M$, an iterative algorithm is provided in \textbf{Algorithm \ref{alg1}} to optimize the input and channel statistics at the same time. Both optimizations are subject to postulates of probability as well as Fano's inequality. Capacity results for accuracy targets $0.99, 9.95, 0.9, 0.85, 0.8$ and $0.75$ are provided in Table 1. Using these results, we can practically determine the upper bounds on the maximum achievable ITR. For instance, if a genuine TI achieves an accuracy of 0.99 using only an observation window of $T=0.2$ seconds, our capacity results imply that the upper bound on ITR can be calculated as $0.9277 \times \frac{60}{0.2} = 278.3$bpm. Such a high ITR is quite untypical in binary symbol communications in the BCI context, and yet it motivates us for a joint design to fully exploit the available capacity. We finally remark that the mapping between the achievable accuracy and the observation window $T$ is also a function of the structure of the data manifold as well as the dimension reduction techniques used before the application of TIs.

\begin{table}[t!]
\small{
\centering
\begin{tabular}{||l||c|c|c|c|c|c||}
\hline
\toprule
\textsf{\textbf{Avg. Accuracy Target}}            & \textsf{\textbf{0.99}}   & \textsf{\textbf{0.95}}   & \textsf{\textbf{0.9}}    & \textsf{\textbf{0.85}}   & \textsf{\textbf{0.8}}    & \textsf{\textbf{0.75}}  \\ \hline \midrule 
\textsf{\textbf{Capacity}}            & 0.9277 & 0.746  & 0.5787 & 0.4412 & 0.3219 & 0.2155 \\ \midrule
\textsf{\textbf{Conditional Entropy}} & 0.0703 & 0.2271 & 0.3367 & 0.3908 & 0.4001 & 0.3655 \\ \midrule
\textsf{\textbf{Fano's Bound}}   & 0.0932 & 0.3627 & 0.6343 & 0.8644 & 1.0613 & 1.2276 \\ \bottomrule
\hline
\end{tabular}
\label{table1}
\caption{Optimization of the channel statistics ($M=2$) to maximize the mutual information (capacity) given the target average accuracy (classification error) rate.}
} 
\end{table}

On the other hand, for $M>2$,  such optimizations can be carried out  based on the separability of the input data using a multi-class classification algorithm. However, as the number of classes increase, the possibility of hitting a local minimum grows, making the outcome unstable and vary at each run of the proposed algorithm (line 3, \textbf{Algorithm 1}). However, multi-class classification can be implemented as an ensemble of multiple binary (weak and unstable) classifications (such as one-vs-one, one-vs-all \cite{galar2011} or expressed in a more general framework called error correction output codes \cite{guney2019}), thus making the arguments of the previous paragraph directly applicable. 

Note that channel parameter optimizations are indeed not the characteristic of telecommunication systems (regarding Shannon's channel coding theorem \cite{shannon1948}). The channel model is usually a given quantity defined by the transmission medium and capacity-achieving IDs are found via solving an optimization problem so as to determine the maximum information transfer rate over the communication medium. Therefore, the practice of this paper will hopefully help us understand what is achievable using TI algorithms or classification techniques subject to a target accuracy threshold. Such an upper bound on the performance will also help compare the performances of different TI algorithms under equal settings and highlight how much improvement they can bring into the ITR enhancements for future BCI systems.  


\textcolor{black}{As a result, we would like to highlight that most offline BCI signal detection techniques employ subject-specific optimizations, such as training the TI based on individual template signals for each observation window. Furthermore, we have observed that the experimental design and the resulting channel formation are closely intertwined, resulting in highly collaborative systems with interconnected components. In this regard, we have recognized that extending the optimizations to include subject-level stimuli design and constructing the requisite set of input patterns for a better channel design could significantly enhance the experienced ITR, thereby narrowing down the considerable performance gaps that exist in published TI schemes in the literature. Therefore, the outcomes of our study underscore the criticality of joint design in SSVEP-based BCIs, emphasizing the need for stronger individual calibrations for each user of the system and tighter symbiosis between the neural circuits of the brain and the digital circuits of the computer.}

\section{Conclusion}
In this study, we have considered a more realistic ITR definition without making superfluous assumptions about the channel transition and input statistics and numerically supported it using two well known datasets along with two prominent TI algorithms in an SSVEP-based BCI context. This numerical definition characterizes the communication rate between the computer and the brain as if sending symbols over a discrete, asymmetric, non-stationary memoryless channel. This definition also provided a set of intriguing ways of comparing all previously developed TIs and highlight where they suffer information-theoretically in achieving better or worse ITRs. Our findings also imply that the proposed ITR definition is particularly important for subject-level customizations, enabling a more realistic measurement tool. Moreover, the proposed definition is shown to help with the design of the channel transitions (binary classification use case) and potentially lead to future work for finding upper bounds on TI performances with large alphabet sizes in SSVEP-based BCI settings. 

\section*{Acknowledgments}

This research was supported in part by Scientific and Technological Research Council of Turkey (TUBITAK) under grant number 1059B192100830.

\end{document}